\documentclass[pre,onecolumn]{revtex4-2}
\usepackage{amsmath}
\usepackage{graphicx}

\begin{document}

\title{Two-dimensional solitons in second-harmonic-generating media with
fractional diffraction}
\author{Hidetsugi Sakaguchi$^{1}$ and Boris A. Malomed$^{2,3}$}

\begin{abstract}
We introduce a system of propagation equations for the fundamental-frequency
(FF) and \ second-harmonic (SH) waves in the bulk waveguide with the
effective fractional diffraction and quadratic ($\chi ^{(2)}$) nonlinearity.
The numerical solution produces families of ground-state (zero-vorticity)
two-dimensional solitons in the free space, which are stable in exact
agreement with the Vakhitov-Kolokolov criterion, while vortex solitons are
completely unstable in that case. Mobility of the stable solitons and
inelastic collisions between them are briefly considered too. In the
presence of a harmonic-oscillator (HO) trapping potential, families of
partially stable single- and two-color solitons (SH-only or FF-SH ones,
respectively) are obtained, with zero and nonzero vorticities. The single-
and two-color solitons are linked by a bifurcation which takes place with
the increase of the soliton's power.
\end{abstract}

\maketitle

\address{$^{1}$Interdisciplinary Graduate School of Engineering Sciences, Kyushu
University, Kasuga, Fukuoka 816-8580, Japan}
\address{$^{2}$Department of Physical Electronics, School of Electrical
Engineering, Faculty of Engineering, and Center for Light-Matter
Interaction, Tel Aviv University, Tel Aviv 69978, Israel}
\address{$^{3}$Instituto de Alta Investigaci\'{o}n, Universidad de
Tarapac\'{a}, Casilla 7D, Arica, Chile}

\section{Introduction}

Fractional derivatives have been known since long ago as a formal extension
of the classical calculus. A broadly accepted definition is one of the
\textit{Caputo derivative} of non-integer order $\alpha $ \cite%
{Uchaikin,Caputo},%
\begin{equation}
D_{x}^{\alpha }\psi (x)=\frac{1}{\Gamma \left( 1-\{\alpha \}\right) }%
\int_{0}^{x}\frac{f^{\left( n\right) }(\xi )dx}{\left( x-\xi \right)
^{\left\{ \alpha \right\} }},  \label{cap}
\end{equation}%
where $n\equiv \lbrack \alpha ]+1$, $[\alpha ]$ and $\{\alpha \}$ being the
integer and fractional parts of $\alpha $, $\Gamma $ is the Gamma-function,
and $f^{(n)}$ is the usual derivative. Note that this definition represents
the fractional derivative as the integral (nonlocal) operator, rather than
as a differential one. In physics, fractional differential operators were
introduced as the basis of the Laskin's fractional quantum mechanics for
particles whose classical motion is performed by L\'{e}vy flights \cite%
{Laskin1,Laskin2,Book-Laskin}. While experimental realization of this theory
has not been reported yet, it was proposed by Longhi to emulate the
fractional Schr\"{o}dinger equation in optics, making use of the well-known
similarity of the quantum-mechanical Schr\"{o}dinger equation and the
equation for the paraxial propagation of light. Thus, the action of the
fractional kinetic-energy operator in the quantum theory can be emulated by
means of the effective fractional diffraction in optics \cite{Longhi}. The
proposal was based on using an optical cavity, decomposing the light beam
into transverse Fourier components and passing them through an appropriate
phase shifter, which imposes phase changes corresponding to the expected
effect of the fractional diffraction. Eventually, the Fourier components are
recombined back into a single optical beam. While the action of this setup
is discrete, circulation of the beam in the cavity is approximated by the
continuous fractional Schr\"{o}dinger equation.

The experimental implementation of the emulated fractional diffraction in
optics has not yet been reported either. However, a similar effect, in the
form of effective fractional group-velocity dispersion (GVD) in a fiber
cavity, has been reported in Ref. \cite{Shilong}. It used the decomposition
of the optical pulse in the temporal domain into its spectral components,
which were given phase shifts, emulating the fractional GVD, by passing them
through an appropriate phase plate (hologram).

The realization of the fractional diffraction or dispersion in optical
systems suggests to take into regard intrinsic nonlinearity of the material,
thus introducing various forms of the fractional nonlinear Schr\"{o}dinger
(FNLS) equation. These models give rise to diverse species of fractional
solitons and related phenomena, that include accessible \cite{fNLSE1}, gap
\cite{fNLSE2,fNLSE5,fNLSE8}, and lattice \cite{fNLSE9,fNLSE10,fNLSE12}
solitons, breathers \cite{fNLSE14}, vortices \cite{fNLSE9,fNLSE18}, wave
collapse \cite{fNLSE3}, spontaneous symmetry breaking \cite{fNLSE14,fNLSE16}%
, nonlinear couplers \cite{CFNLSE1,DW-FNLSE,Strunin}, spin-orbit-coupled
systems \cite{CFNLSE3,we}, etc. In addition to these models formulated in
the 1D and 2D spatial domains, fractional spatiotemporal models with the
nonlocal nonlinearity of the \textquotedblleft accessible" type and various
potentials were introduced in terms of spherical coordinates \cite%
{Zhong1,Zhong2,Zhong3}. These theoretical results were reviewed in Refs.
\cite{Malomed-Rev}, \cite{Review2}, and \cite{Review3}. Very recently, the
first experimental realization of nonlinear effects in a fiber cavity with
fractional GVD was reported, in the form of bifurcations of spectral peaks
\cite{arXiv}.

The above-mentioned works considered the interplay of the fractional
diffraction/dispersion with the ubiquitous cubic ($\chi ^{(3)}$) term in the
framework of the FNLS equation. Another generic species of the optical
nonlinearity, of the $\chi ^{(2)}$ type, represents the second-harmonic (SH)
generation by quadratic terms, in the framework of coupled equations for the
fundamental-frequency (FF) and SH waves \cite{Torruellas,Torner,Jena,Buryak}%
. The system of the one-dimensional (1D) FF-SH equations, coupled by $\chi
^{(2)}$ terms, with the fractional diffraction acting on both the FF and SH
components, was recently introduced in Ref. \cite{1D}, where existence and
stability conditions for fundamental solitons generated by this system were
identified by means of numerical calculations.

In this connection, it is relevant to mention that a model based on the
single one-dimensional FNLS equation, combining the cubic self-repulsive and
quadratic attractive terms, was investigated in another recent work \cite%
{fNLSE20}. This combined nonlinearity appears in the effective 1D
Gross-Pitaevskii (GP) equation for identical wave functions of two
components of the binary Bose-Einstein condensate (BEC), which takes into
regard the effective repulsive mean-field interaction and the correction
produced by effects of quantum fluctuations \cite{PA}. The quadratic-cubic
fractional GP equation introduced in Ref. \cite{fNLSE20} is, presumably, a
relevant model for the quasi-1D BEC of particles which move by L\'{e}vy
flights, obeying the above-mentioned fractional Schr\"{o}dinger equation in
the single-particle limit. Soliton families, produced by this fractional GP
equation, and their stability were studied in Ref. \cite{fNLSE20}, including
bound states of fundamental solitons supported by the lattice modulation of
the quadratic term.

The objective of the present work is to introduce a two-dimensional (2D)
system of equations for the FF and SH waves with the optical $\chi ^{\left(
2\right) }$ nonlinearity and fractional diffraction acting in both
equations. The consideration of this system \ is definitely relevant -- in
particular, because stable spatial optical solitons supported by the $\chi
^{(2)}$ nonlinearity (in the case of the normal non-fractional diffraction)
were first created in the (2+1)D form \cite{Torruellas}. We consider the
system in the free space, and separately in the presence of a 2D
harmonic-oscillator (HO) trapping potential. Using numerical methods, we
construct families of 2D ground-state (GS, alias zero-vorticity) and vortex
solitons and identify their stability. In the free space, the stability of
the GS solitons exactly obeys the celebrated Vakhitov-Kolokolov (VK )
criterion \cite{VK,Berge} (its precise form is defined below), while the
vortex solitons are unstable, as may be expected from the comparison with
known results for vortex solitons in the case of the non-fractional
diffraction \cite{Skryabin,Petrov}. The HO potential may provide
stabilization for various modes -- in particular, for the \textquotedblleft
single-color" bound states which contain solely the SH component, while the
FF one is zero, as well as for the \textquotedblleft two-color" modes
including both components. In spite of the apparent simplicity of the former
states, their stability is not obvious, as it is affected by the $\chi ^{(2)}
$ coupling of small perturbations in the FF component to the underlying SH
field, which may give rise to the parametric instability \cite{Hidetsugu}.
Furthermore, we demonstrate that the trapping potential provides partial
stabilization for vortex states as well, of both the single- and two-color
types.

The paper is organized as follows. The model is introduced in Section \ref%
{Sec II}, which is followed in Section \ref{Sec III} by systematic
presentation of results for soliton families and their stability in the free
space (as mentioned above, the stability of the GS solitons fully agrees
with the Vakhitov-Kolokolov criterion). Section \ref{Sec III} also briefly
examines mobility of stable fundamental solitons and collisions between
them. The results for single- and two-color fundamental and vortex solitons
supported by the HO potential are reported in Section \ref{Sec IV}. The
paper is concluded by Section \ref{Sec V}.

\section{The model}

\label{Sec II}

The system of coupled 2D equations for the spatial-domain evolution of
amplitudes $U\left( x,y,z\right) $ and $W\left( x,y,z\right) $ of the FF and
SH waves in the bulk optical waveguide with fractional diffraction is
\begin{eqnarray}
i\frac{\partial U}{\partial z} &=&\frac{1}{2}\left( -\frac{\partial ^{2}}{%
\partial x^{2}}-\frac{\partial ^{2}}{\partial y^{2}}\right) ^{\alpha
/2}U-WU^{\ast },  \label{U} \\
2i\frac{\partial W}{\partial z} &=&\frac{1}{2}\left( -\frac{\partial ^{2}}{%
\partial x^{2}}-\frac{\partial ^{2}}{\partial y^{2}}\right) ^{\alpha /2}W+QW-%
\frac{1}{2}U^{2}.  \label{W}
\end{eqnarray}%
Here $\left( x,y\right) $ are the transverse coordinates, $z$ is the
propagation distance, real $Q$ is the mismatch parameter, which can be
reduced, by rescaling, to one of three values,
\begin{equation}
Q=\pm 1\text{ \textrm{or~}}Q=0,  \label{QQ}
\end{equation}%
and $\ast $ stands for the complex conjugate.

The present system is introduced as a natural 2D extension of the 1D
fractional $\chi ^{(2)}$ system that was elaborated in Ref. \cite{1D}.
According to the above-mentioned derivation \cite{Longhi}, the
fractional-diffraction operator is represented by the 2D version of the
\textit{Riesz derivative} \cite{Riesz,Riesz2} with \textit{L\'{e}vy index}
(LI) $\alpha $, \textit{viz}.,
\begin{equation}
\left( -\frac{\partial ^{2}}{\partial x^{2}}-\frac{\partial ^{2}}{\partial
y^{2}}\right) ^{\alpha /2}U(x)=\frac{1}{\left( 2\pi \right) ^{2}}%
\int_{-\infty }^{+\infty }dp\int_{-\infty }^{+\infty }dq\left(
p^{2}+q^{2}\right) ^{\alpha /2}\int_{-\infty }^{+\infty }d\xi \int_{-\infty
}^{+\infty }d\eta e^{ip(x-\xi )+iq(y-\eta )}U(\xi ,\eta ).
\label{Risz derivative}
\end{equation}%
The usual and fractional diffraction corresponds, respectively, to $\alpha
=2 $ and $\alpha <2$. A straightforward estimate demonstrates that the $\chi
^{(2)}$ system of 2D equations (\ref{U}) and (\ref{W}) gives rise to the
collapse (catastrophic self-compression leading to appearance of
singularities \cite{Berge}) at $\alpha \leq 1$, hence the system may give
rise to stable solitons in the interval of
\begin{equation}
1<\alpha \leq 2  \label{no-coll}
\end{equation}%
(the 1D version of the fractional $\chi ^{(2)}$ system supports stable
soliton in the interval of $1/2<\alpha \leq 2$). On the other hand, the 2D
FNLS equation with the cubic self-focusing term gives rise to the collapse
at all values of $\alpha \leq 2$, hence it cannot support stable solitons in
any case \cite{fNLSE3,Malomed-Rev}).

It is relevant to stress that the conclusion of the collapse avoidance in
interval (\ref{no-coll}) is valid for the free space. If a trapping
potential is added to Eqs. (\ref{U}) and (\ref{W}). stability of the
localized modes may be extended to the region of $\alpha \leq 1$, as
demonstrated below.

Stationary solutions to Eqs. (\ref{U}) and (\ref{W}) with FF and SH
propagation constants $k>0$ and $2k$ are looked for as%
\begin{equation}
U\left( x,y,z\right) =e^{ikz}u(x,y),W\left( x,y,z\right) =e^{2ikz}w(x,y),
\label{UW}
\end{equation}%
where functions $u$ and $w$ satisfy the following system of stationary
equations:
\begin{eqnarray}
-ku &=&\frac{1}{2}\left( -\frac{\partial ^{2}}{\partial x^{2}}-\frac{%
\partial ^{2}}{\partial y^{2}}\right) ^{\alpha /2}u-wu^{\ast },  \label{u} \\
-4kw &=&\frac{1}{2}\left( -\frac{\partial ^{2}}{\partial x^{2}}-\frac{%
\partial ^{2}}{\partial y^{2}}\right) ^{\alpha /2}w+Qw-\frac{1}{2}u^{2}.
\label{w}
\end{eqnarray}%
Stationary localized solutions are naturally characterized by the total
power, which is a dynamical invariant of Eqs. (\ref{U}) and (\ref{W}):
\begin{equation}
P=\int_{-\infty }^{+\infty }dx\int_{-\infty }^{+\infty }dy(|u|^{2}+4|w|^{2}).
\label{P}
\end{equation}

Axisymmetric stationary solutions of Eqs. (\ref{u}) and (\ref{w}) are sought
for, in terms of polar coordinates $\left( r,\theta \right) $, as
\begin{equation}
u\left( r,\theta \right) =u_{m}(r)e^{im\theta },~w\left( r,\theta \right)
=w_{m}(r)e^{2im\theta },  \label{m}
\end{equation}%
where $u_{m}(r)$ and $w_{m}(r)$ are real radial functions, and integer $%
m=0,\pm 1,\pm 2\,...$ is the vorticity (alias the winding number) of the FF
component. Then, the SH\ winding number is $2m$, as determined by Eq. (\ref%
{w}). In addition to the power (\ref{P}), the stationary states are
characterized by the angular momentum,%
\begin{equation}
M=2\pi m^{2}\int_{0}^{\infty }rdr\left[ u_{m}^{2}(r)+8w_{m}^{2}(r)\right] ,
\label{M}
\end{equation}%
which also represents a dynamical invariant of the underlying system.

Lastly, the system's Hamiltonian, which is conserved too, is written in
terms of stationary fields $u$ and $w$ as%
\begin{gather}
H=\frac{1}{\left( 2\pi \right) ^{2}}\int_{0}^{+\infty }dp\int_{0}^{+\infty
}dq\left( p^{2}+q^{2}\right) ^{\alpha /2}\int_{-\infty }^{+\infty
}dx\int_{-\infty }^{+\infty }dy\int_{-\infty }^{+\infty }d\xi \int_{-\infty
}^{+\infty }d\eta  \notag \\
\times \cos \left[ p\left( x-\xi \right) \right] \cos \left[ q\left( y-\eta
\right) \right] \left[ u^{\ast }\left( x,y\right) u\left( \xi ,\eta \right)
+w^{\ast }\left( x,y\right) w\left( \xi ,\eta \right) \right]  \notag \\
+\int_{-\infty }^{+\infty }dx\int_{-\infty }^{+\infty }dy\left\{ \left[
\frac{Q}{2}\left\vert w\right\vert ^{2}\right] -\frac{1}{4}\left[ w\left(
u^{\ast }\right) ^{2}+w^{\ast }u^{2}\right] \right\} ,  \label{H}
\end{gather}%
where the complex structure of the fractional-diffraction terms corresponds
to the definition (\ref{Risz derivative}). Note that this structure secures
that $H$ is always real.

In the case of zero mismatch, $Q=0$ (see Eq. (\ref{QQ})), it follows from
Eqs. (\ref{u}) and (\ref{w}) that amplitudes $U_{0}=u_{0}(r=0)$ and $%
W_{0}=w_{0}(r=0)$ of real solutions for GS axisymmetric solitons (\ref{m})
with $m=0$, their characteristic radial size $R$, and the corresponding
power (\ref{P}) satisfy exact scaling relations for the variation of $k$:%
\begin{equation}
U_{0}(Q=0),W_{0}\left( Q=0\right) \sim k,~R\left( Q=0\right) \sim
k^{-1/\alpha },~P(Q=0)\sim k^{2(1-1/\alpha )}.  \label{scaling Q=0}
\end{equation}%
In the presence of the mismatch, i.e., with $Q=\pm 1$ in Eq. (\ref{QQ}), the
scaling relations (\ref{scaling Q=0}) are asymptotically valid for $k\gg 1$,
i.e., for narrow solitons with $R\ll 1$, as confirmed below by Fig. \ref%
{fig2}(b).

In the interval of LI values (\ref{no-coll}), relation $P\left( k\right) $,
as given by Eq. (\ref{scaling Q=0}), satisfies the above-mentioned VK
criterion, which takes the form of $dP/dk>0$. It is the universal necessary
stability condition for localized modes maintained by any self-attractive
nonlinearity. The fact that the $\chi ^{(2)}$ nonlinearity may be considered
as the attractive self-interaction follows from the fact that, in the GS
states produced by Eqs. (\ref{U}) and (\ref{W}), the real $w$ component is
\emph{positive} (see Fig. \ref{fig1}(a) below), hence the $\chi ^{(2)}$ term
in Hamiltonian (\ref{H}) is \emph{negative}, which indeed implies the
self-attraction.

Precisely at $\alpha =1$, Eq. (\ref{scaling Q=0}) yields $dP/dk=0$, implying
the occurrence of the \textit{critical collapse}, which takes place in the
framework of the system of equations (\ref{U}) and (\ref{W}) if the power of
the initial condition exceeds a certain critical value \cite{Berge}. At $%
\alpha <1$, Eq. (\ref{scaling Q=0}) yields $dP/dk<0$, i.e., the solitons
disobey the VK criterion. This implies the onset of the \textit{%
supercritical collapse}, which may be initiated by the input with an
arbitrarily small power (in other words, the respective critical value of
the power is zero \cite{Berge}). An example of the collapse is demonstrated
below in Fig. \ref{fig7}.

\section{Solitons and their stability in the free space}

\label{Sec III}

\subsection{Fundamental solitons}

\begin{figure}[h]
\begin{center}
\includegraphics[height=4.cm]{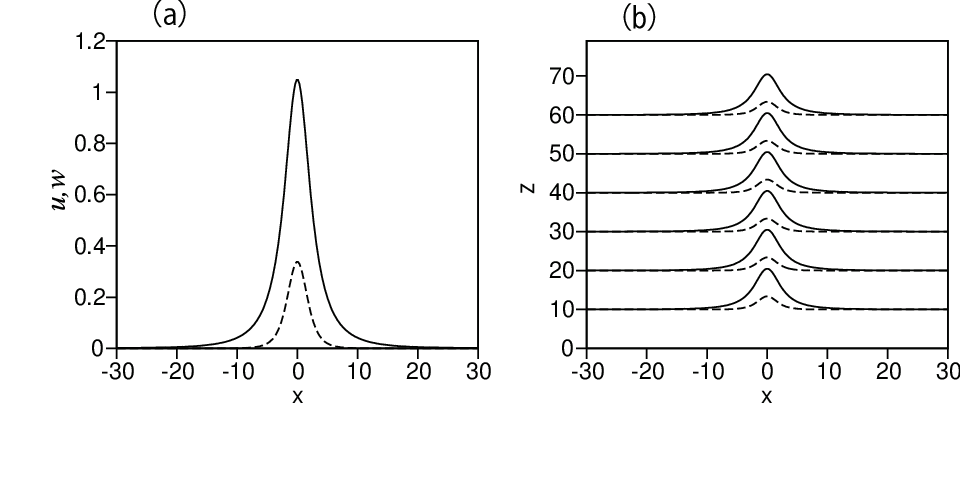}
\end{center}
\caption{(a) Profiles $u(x)$ and $w(x)$ (solid and dashed lines,
respectively) of the cross section, drawn through $y=0$, of the FF and SH
components of the ground-state (zero-vorticity) stationary soliton solution
of Eqs. (\protect\ref{U}) and (\protect\ref{W}) with LI $\protect\alpha =1.5$%
, $k=0.0583$, and mismatch $Q=1$. It is tantamount to the solution of Eqs. (%
\protect\ref{u}) and (\protect\ref{w}) with the same parameters. The total
power (\protect\ref{P}) of the soliton is $P=24.5$. (b) Snapshots of $|U(x)|$
and $|W(x)|$, taken at $z=10n$ ($n=1,2,\cdots ,7)$, corroborate the stable
evolution of this perturbed soliton.}
\label{fig1}
\end{figure}

Numerical solutions for stationary fundamental solitons were produced by
means of the imaginary-time (IT) simulation method \cite{Bao,Yang} applied
to Eqs. (\ref{U}) and (\ref{W}) (in the present notation, it actually means
the evolution in imaginary $z$). The initial input is a Gaussian function
for $U$ and $W=0$. The stability of the so produced stationary states was
verified by direct real-time (actually, real-$z$)\_simulations of their
perturbed evolution. A typical example of a stable 2D soliton, obtained for $%
Q=+1$ in Eq. (\ref{QQ}), is plotted, by means of its cross sections, in Fig. %
\ref{fig1}.

\begin{figure}[h]
\begin{center}
\includegraphics[height=4.cm]{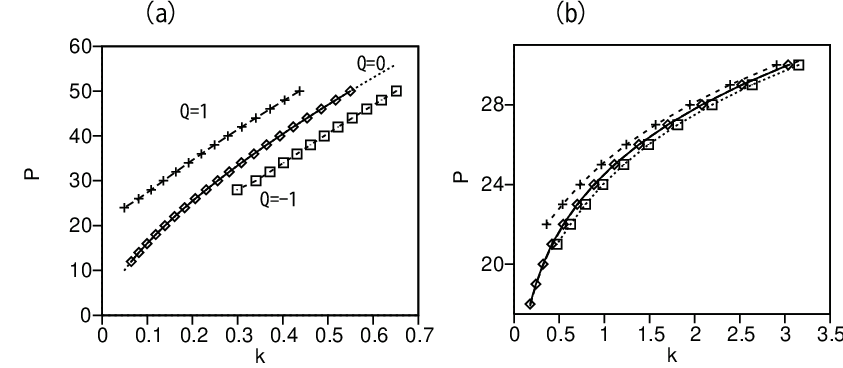}
\end{center}
\caption{The relation between the total power (\protect\ref{P}) of stable
fundamental solitons and their FF propagation constant $k$, at LI values $%
\protect\alpha =1.5$ (a) and $\protect\alpha =1.1$ (b), for the three fixed
values of the mismatch, $Q=+1$ (crosses), $0$ (rhombuses), and $-1$
(squares), see Eq. (\protect\ref{QQ}).}
\label{fig2}
\end{figure}

In Fig. \ref{fig2}, families of fundamental solitons are characterized by
the respective $P(k)$ dependences for two values of LI: an intermediate one,
$\alpha =1.5$, and $\alpha =1.1$, which is close to the collapse threshold ($%
\alpha =1$), in terms of interval (\ref{QQ}). The dependences are plotted
for the three values of the mismatch, $Q=+1$, $0$, and $-1$, as fixed in Eq.
(\ref{QQ}). All the families are stable, in accordance with the fact that
they satisfy the VK criterion, $dP/dk>0$, and in agreement with the
well-known stability of the similar fundamental 2D\ solitons in the $\chi
^{(2)}$ system with the normal (non-fractional) diffraction, $\alpha =2$.
The dotted lines in Fig. \ref{fig2}(a) and (b) are $P=74.5k^{2/3}$ and $%
P=24.5k^{0.2/1.1}$, respectively, with the powers predicted by Eq. (\ref%
{scaling Q=0}), and fitting numerical factors in front of the powers. These
curves confirm the agreement of the numerical findings for $Q=0$ with the
analytical prediction.

\begin{figure}[h]
\begin{center}
\includegraphics[height=4.cm]{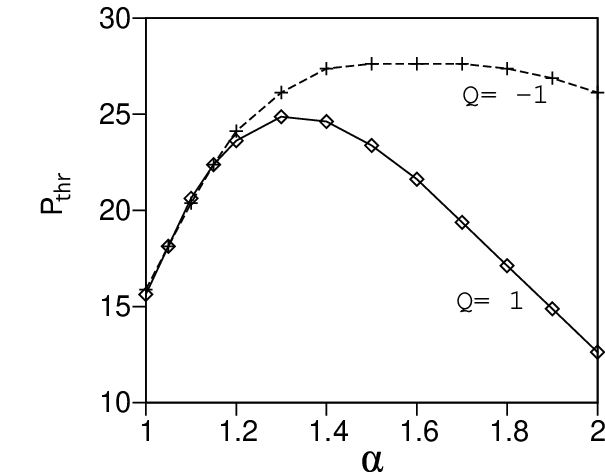}
\end{center}
\caption{The threshold values of the power below which stable solitons do
not exist at $Q=+1$ and $-1$( the solid and dashed lines, respectively).}
\label{fig3}
\end{figure}

Also in agreement with the analytical prediction for $Q=0$, the GS solitons
exist and are stable at all values of $k$ (the same is found for other
values of LI from interval (\ref{QQ})). On the other hand, for $Q=\pm 1$,
the soliton solutions exist above the threshold (minimum) value of the
power, $P\geq P_{\mathrm{thr}}$, and, accordingly, at $k\geq k_{\mathrm{thr}%
} $. The threshold power is plotted, as a function of LI, in Fig. \ref{fig3}%
. The comparison with the 1D version of the $\chi ^{(2)}$ system, which was
explored in Ref. \cite{1D}, suggests that, at $Q=\pm 1$ and $k<k_{\mathrm{thr%
}}$, there should exist another, strongly unstable, branch of the soliton
solutions (also taking values $P\geq P_{\mathrm{thr}}$), which merges with
the stable one at $k=k_{\mathrm{thr}}$ and $P=P_{\mathrm{thr}}$. However,
the IT simulations do not converge to strongly unstable solutions (in Ref.
\cite{1D}, the 1D unstable solutions were found by means of the Newton
conjugate gradient method \cite{Yang}, which is not sensitive to the
stability of the stationary solutions). On the other hand, at $P<P_{\mathrm{%
thr}}$ the IT simulations converge to a spatially uniform state .

\begin{figure}[h]
\begin{center}
\includegraphics[height=4.cm]{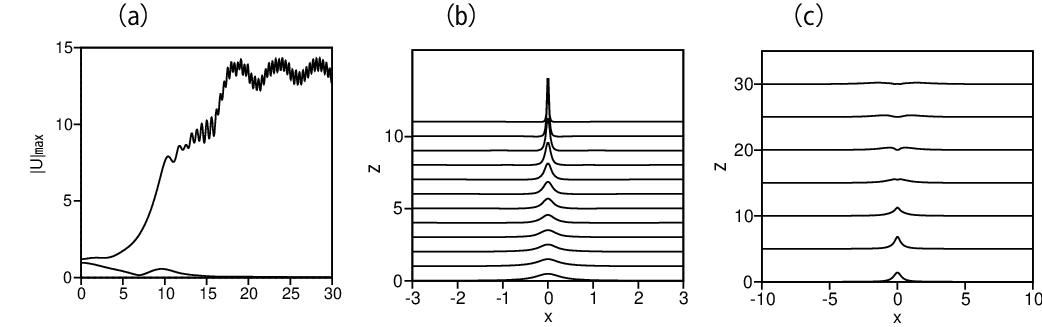}
\end{center}
\par
.
\caption{(a) The evolution of the FF amplitude, $|U(x=0,y=0,z)|$, as
produced by simulations of Eqs. (\protect\ref{U}) and (\protect\ref{W}) with
$\protect\alpha =0.4$ and $Q=0$, and the initial conditions taken as per Eq.
(\protect\ref{input1}) (the collapsing solution) or Eq. (\protect\ref{input2}%
) (the decaying one). (b) and (c): The corresponding evolution of the cross
section of the FF profile, $|U(x,y=0,z)|$}
\label{fig7}
\end{figure}

The expected onset of the critical and supercritical collapse in the system
of Eqs. (\ref{U}) and (\ref{W}) at $\alpha =1$ and $\alpha <1$,
respectively, was also tested in direct simulations. As an example, Fig. \ref%
{fig7} presents the findings for $\alpha =0.4$ (the deeply supercritical
case) and $Q=0$, produced by inputs which were taken at $z=0$ as%
\begin{eqnarray}
U(x,y,0) &=&(1.21+0.88i)/\{9(x^{2}+y^{2})+1\},~W(x,y,0)=0;  \label{input1} \\
U(x,y,0) &=&(0.99+0.72i)/\{9(x^{2}+y^{2})+1\},~W(x,y,0)=0.  \label{input2}
\end{eqnarray}%
Figure \ref{fig7}(a) shows that the former input gives rise to onset of the
collapse through the steep growth of amplitude $\left\vert U\left(
x=0,y=0,z\right) \right\vert $. On the other hand, input (\ref{input2}),
with a somewhat smaller power, initiates the evolution which leads to the
decay of the localized configuration (in the case which\ is generically
categorized as the collapse, decay of particular inputs is possible too).
The development of the collapse and decay, which are initiated, severally,
by inputs (\ref{input1}) and (\ref{input2}) is directly illustrated by the
evolution of the respective cross-section profiles in Figs. \ref{fig7}(b)
and (c), respectively.

\subsection{Moving (tilted) solitons}

The fractional diffraction operators destroy the Galilean invariance of Eqs.
(\ref{U}) and (\ref{W}), therefore producing moving solitons (actually, ones
tilted in the spatial domain) is a nontrivial problem. For this purpose, the
underlying equation are rewritten in the tilted coordinates, with $%
x\rightarrow \tilde{x}\equiv x-vz$, where $v$ is the \textquotedblleft
velocity" (in fact, the tilt in the $\left( x,z\right) $ plane):%
\begin{eqnarray}
i\frac{\partial U}{\partial z}-iv\frac{\partial U}{\partial \tilde{x}} &=&%
\frac{1}{2}\left( -\frac{\partial ^{2}}{\partial \tilde{x}^{2}}-\frac{%
\partial ^{2}}{\partial y^{2}}\right) ^{\alpha /2}U-WU^{\ast },  \label{Uv}
\\
2i\frac{\partial W}{\partial z}-2iv\frac{\partial W}{\partial \tilde{x}} &=&%
\frac{1}{2}\left( -\frac{\partial ^{2}}{\partial \tilde{x}^{2}}-\frac{%
\partial ^{2}}{\partial y^{2}}\right) ^{\alpha /2}W+QW-\frac{1}{2}U^{2}.
\label{Wv}
\end{eqnarray}

\begin{figure}[h]
\begin{center}
\includegraphics[height=4.cm]{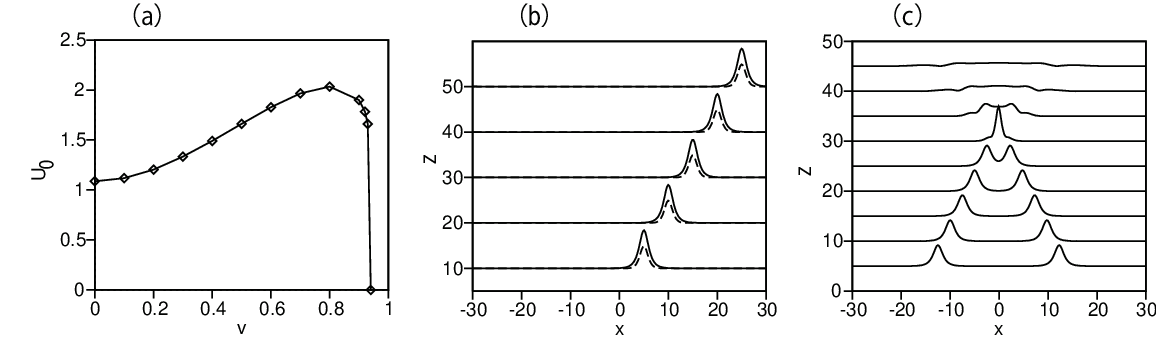}
\end{center}
\caption{(a) The amplitude $U_{0}$ of the FF component of the
\textquotedblleft moving" (tilted) solitons with fixed power $P=20$ vs. the
\textquotedblleft velocity" $v$, as produced by the IT method applied to
Eqs. (\protect\ref{Uv}) and (\protect\ref{Wv}) with $\protect\alpha =1.5$, $%
Q=0$. (b) The evolution of $|U(x,z)|$ (the solid line) and $|W(x,z)|$ (the
dashed line) for a stable \textquotedblleft moving" (tilted) soliton with $%
v=0.5$, displayed in the cross section $y=0$. (c) The inelastic collision,
leading to mutual destruction of the two moving solitons with $v=\pm 0.5$.}
\label{fig4}
\end{figure}

Stationary solutions for tilted solitons were produced by means of the IT
algorithm applied to Eqs. (\ref{Uv}) and (\ref{Wv}), and their stability was
again tested by simulations of the perturbed evolution in real time (real $z$%
, in fact). A typical family of stable soliton solutions is represented in
Fig. \ref{fig4}(a) by the dependence of the FF amplitude, $U_{0}=\left\vert
U\left( \tilde{x}=0,y=0\right) \right\vert $ vs. the tilt $v$ for a fixed
value of the total power, $P=20$, setting $\alpha =1.5$ and $Q=0$ in Eqs. (%
\ref{Uv}) and (\ref{Wv}). It is seen that the solitons gradually shrink,
increasing the amplitude, as $v$ grows from $0$ to $0.8$, and then rapidly
expand, featuring steep fall of the amplitude. The soliton solution ceases
to exist through delocalization at $v\approx 0.94$, when its amplitude
vanishes. At $v>0.94$, the IT integration converges to a flat state. Figure %
\ref{fig4}(b) demonstrates an example of the stable soliton tilted with
slope $v$ $=0.5$.

Of course, the solitons also exist with the opposite velocity (tilt), $-v,$
which makes it possible to simulate collisions between them. Figure \ref%
{fig4}(c) demonstrates that the collisions are fully inelastic, leading to
mutual destruction of the solitons.

\subsection{Unstable vortex solitons}

\begin{figure}[h]
\begin{center}
\includegraphics[height=4.cm]{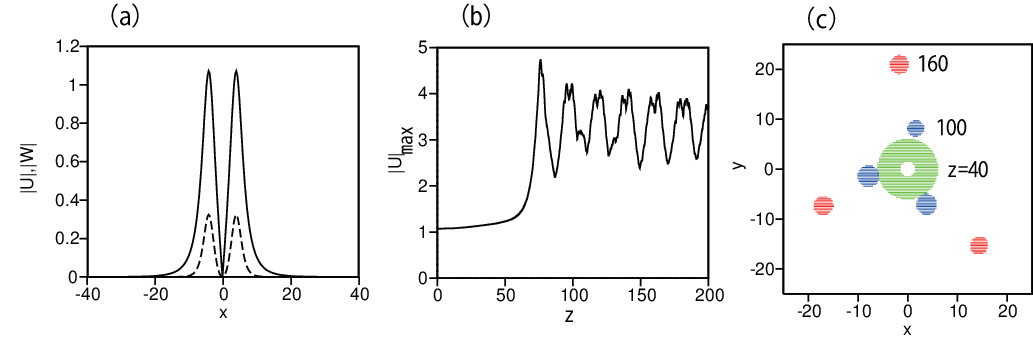}
\end{center}
\caption{(a) Profiles of $|U(x)|$ and $|W(x)|$ in cross section $y=0$ for
the vortex soliton with $m=1$ and power $P=130$, obtained as a stationary
numerical solution for $\protect\alpha =1.5$ and $Q=+1$ in Eqs. (\protect\ref%
{u}) and (\protect\ref{w}). (b) The maximum value of the FF component, $%
|U\left( x,y\right) |$, of the same soliton vs. propagation distance $z$,
demonstrating the soliton's instability. (c) For the same unstable soliton,
the snapshots of the high-field regions, with $|U(x,y)|>0.5$, at $z=40$
(green), $100$ (blue), and $160$ (red), demonstrate that the original vortex
ring spontaneously splits into the set of three moving fundamental
(zero-vorticity) solitons in an excited state (with internal vibrations).}
\label{fig5}
\end{figure}

\begin{figure}[h]
\begin{center}
\includegraphics[height=4.cm]{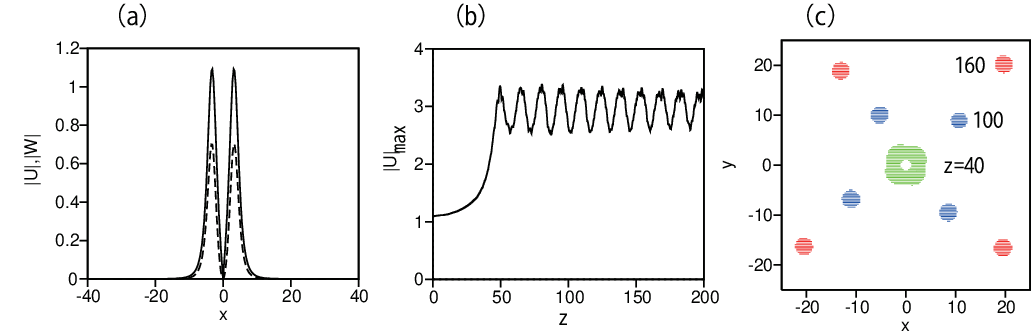}
\end{center}
\caption{The same as in Fig. \protect\ref{fig5}, but for the unstable vortex
soliton with $P=150$, in the case of $Q=-1$. In this case, panel (c)
demonstrates spontaneous fission into a set of four fundamental solitons in
the excited form..}
\label{fig6}
\end{figure}

As well as the usual $\chi ^{(2)}$ system with the normal diffraction ($%
\alpha =2$), the present one, with LI $\alpha <2$, admits the existence of
solitons with embedded vorticity, defined as per Eq. (\ref{m}), with integer
winding numbers $m$ and $2m$ in the phases of the FF and SH components,
respectively. In particular, starting with the input $U\sim e^{i\theta }$
and $W\sim e^{2i\theta }$, the IT method, applied to Eqs. (\ref{U}) and (\ref%
{W}) with $\alpha =1.5$, produces a family of stationary soliton solutions
in the form of vortex rings, with $m=1$ and powers $P\geq 111$ or $P\geq 139$
for $Q=+1$ and $-1$, respectively. These findings are illustrated, in Figs. %
\ref{fig5}(a) and \ref{fig6}(a), by plots of the cross sections of the FF
and SH components of the vortex soliton (ring) for $P=150$ and mismatch
parameters $Q=+1$ and $-1$, respectively.

Similar to the usual 2D $\chi ^{(2)}$ solitons \cite{Skryabin,Petrov}, the
ones produced by Eqs. (\ref{U}) and (\ref{W}) are subject to instability
against spontaneous fission into fragments, which are close to individual
stable fundamental solitons performing intrinsic vibrations, as shown in
Figs. \ref{fig5}(c) and \ref{fig6}(c). The secondary solitons produced by
the fission of the vortex ring are moving, to provide the conservation of
the angular momentum (see Eq. (\ref{M})). It is seen that the fission leads
to effective shrinkage of the fragments, in comparison to the original
vortex ring. In \ agreement with this observation, Figs. \ref{fig5}(b) and %
\ref{fig6}(b) demonstrate that, in the course of the fission, the amplitude
(largest value of the FF component, $\left\vert U(x,y)\right\vert $)
increases to values which are essentially higher than the initial one, and
then performs residual oscillations. The latter feature implies that, as
said above, the fundamental solitons are produced by the fission in an
excited state (with internal vibrations). A similar fission of a vortex
soliton into three fragments was observed at $\alpha =1.5$ and $P=140$ for $%
Q=0$.

\section{Solitons supported by the trapping potential and their stability}

\label{Sec IV}

\subsection{Basic equations}

The confining structure of the bulk waveguide may be represented by a
trapping potential $V\left( x,y\right) $ added to Eqs. (\ref{U}) and (\ref{W}%
), which, in many cases, is taken in the isotropic quadratic form, $%
V(x,y)=V_{0}\left( x^{2}+y^{2}\right) $ with $V_{0}>0$ \cite{Agrawal}. This
modification of the $\chi ^{(2)}$ system is relevant as the confining
potential may help to stabilize solitons \cite{Hidetsugu}.

Thus, we now address the $\chi ^{(2)}$ system with the quadratic potential:
\begin{eqnarray}
i\frac{\partial U}{\partial z} &=&\frac{1}{2}\left( -\frac{\partial ^{2}}{%
\partial x^{2}}-\frac{\partial ^{2}}{\partial y^{2}}\right) ^{\alpha
/2}U-WU^{\ast }+\frac{1}{8}\left( x^{2}+y^{2}\right) U,  \label{potU} \\
2i\frac{\partial W}{\partial z} &=&\frac{1}{2}\left( -\frac{\partial ^{2}}{%
\partial x^{2}}-\frac{\partial ^{2}}{\partial y^{2}}\right) ^{\alpha /2}W+QW-%
\frac{1}{2}U^{2}+\frac{1}{2}\left( x^{2}+y^{2}\right) W.  \label{potW}
\end{eqnarray}%
Here, the coefficient in front of the potential term in Eq. (\ref{potU}) is
fixed to be $1/8$ by means of scaling, and the fact that its counterpart in
Eq. (\ref{potW}) is larger by the factor of $4$ is a generic feature of the $%
\chi ^{(2)}$ system \cite{Buryak}. Stationary solutions of Eqs. (\ref{potU})
and (\ref{potW}) are looked for as per Eq. (\ref{UW}) with the FF
propagation constant $k$, and functions $u$ and $w$ satisfying the equations
\begin{eqnarray}
-ku &=&\frac{1}{2}\left( -\frac{\partial ^{2}}{\partial x^{2}}-\frac{%
\partial ^{2}}{\partial y^{2}}\right) ^{\alpha /2}u-wu^{\ast }+\frac{1}{8}%
\left( x^{2}+y^{2}\right) u,  \label{potu} \\
-4kw &=&\frac{1}{2}\left( -\frac{\partial ^{2}}{\partial x^{2}}-\frac{%
\partial ^{2}}{\partial y^{2}}\right) ^{\alpha /2}w+Qw-\frac{1}{2}u^{2}+%
\frac{1}{2}\left( x^{2}+y^{2}\right) w.  \label{potw}
\end{eqnarray}

\subsection{Soliton solutions}

\subsubsection{Single-color states}

First of all, Eqs. (\ref{potU}) and (\ref{potW}) admit obvious trapped
SH-only (\textit{single-color}) states with $U=0$, while $W$ is, obviously,
tantamount to eigenstates of the linear fractional Schr\"{o}dinger equation
with the HO potential:
\begin{equation}
2i\frac{\partial W}{\partial z}=\frac{1}{2}\left( -\frac{\partial ^{2}}{%
\partial x^{2}}-\frac{\partial ^{2}}{\partial y^{2}}\right) ^{\alpha /2}W+QW+%
\frac{1}{2}\left( x^{2}+y^{2}\right) W.  \label{linear}
\end{equation}%
Stationary solutions to Eq. (\ref{linear}) are looked for in the usual form,
$W\left( x,y,z\right) =e^{2ikz}w(x,y)$ (see Eq. (\ref{UW})), with function $%
w(x,y)$ satisfying the equation%
\begin{equation}
-4kw=\frac{1}{2}\left( -\frac{\partial ^{2}}{\partial x^{2}}-\frac{\partial
^{2}}{\partial y^{2}}\right) ^{\alpha /2}w+Qw+\frac{1}{2}\left(
x^{2}+y^{2}\right) w,  \label{linear-w}
\end{equation}
cf. Eq. (\ref{potw}).

The GS corresponds to the solution of Eq. (\ref{linear-w}) with zero SH
vorticity, $m_{2}=0$, while $m_{2}=1,2,...$ correspond to excited states. In
terms of the general form of the two-color vortex modes (\ref{m}), we have $%
m_{2}\equiv 2m$, hence odd values of $m_{2}$ may represent solely
single-color modes.

\paragraph{The single-color state in the finite-size system in the limit of
vanishing LI, $\protect\alpha \rightarrow 0$}

First, we consider the case of $\alpha \rightarrow 0$.
In terms of the split-step Fourier method implementing the numerical
simulations, term $(-\partial ^{2}/\partial x^{2}-\partial ^{2}/\partial
y^{2})^{\alpha /2}w$ in Eq. (\ref{linear-w}) is calculated by means of the
Fourier transform as $\sum_{p_{x},p_{y}}(p_{x}^{2}+p_{y}^{2})^{\alpha
/2}w_{p_{x},p_{y}}e^{ip_{x}x+ip_{y}y}$, where, in the limit if $\alpha
\rightarrow 0$, we set $(p_{x}^{2}+p_{y}^{2})^{\alpha /2}=1$ for $%
p_{x}^{2}+p_{y}^{2}\neq 0$, and $(p_{x}^{2}+p_{y}^{2})^{\alpha /2}=0$ for $%
p_{x}=p_{y}=0$. Thus, because the dc (constant) Fourier component,
corresponding to $p_{x}=p_{y}=0$, is eliminated, the spatial average of $%
(-\partial ^{2}/\partial x^{2}-\partial ^{2}/\partial y^{2})^{\alpha /2}w$
vanishes. Then, $w\left( x,y\right) $ is recovered by the inverse Fourier
transform, with the first term on the right-hand side of Eq.~(\ref{linear-w}%
) taking the form of
\begin{equation}
\frac{1}{2}w\left( x,y\right) -\frac{1}{2L^{2}}\int_{-L/2}^{+L/2}%
\int_{-L/2}^{+L/2}w\left( x^{\prime },y^{\prime }\right) dx^{\prime
}dy^{\prime },  \label{alpha=0}
\end{equation}%
where $L\times L$ is the actual system's size, and expression (\ref{alpha=0}%
) secures the vanishing of the dc value, as said above.

In this case, we attempt to approximate the GS solution of Eq. (\ref%
{linear-w}) in the finite domain by a Lorentzian,
\begin{equation}
w=\frac{w_{0}}{1+\gamma (x^{2}+y^{2})}.  \label{Lorentz}
\end{equation}%
The substitution of this in Eq.~(\ref{linear-w}) yields
\begin{equation}
\frac{-4kw_{0}}{1+\gamma (x^{2}+y^{2})}=\frac{1}{2}\left[ \frac{w_{0}}{%
1+\gamma (x^{2}+y^{2})}-\bar{w}\right] +Q\frac{w_{0}}{1+\gamma (x^{2}+y^{2})}%
+\frac{x^{2}+y^{2}}{2}\frac{w_{0}}{1+\gamma (x^{2}+y^{2})},
\label{substitution}
\end{equation}%
where the spatial average $\bar{w}=(1/L^{2})\int \int w_{0}/\{1+\gamma
(x^{2}+y^{2})\}dxdy$ in the finite domain is approximated as
\begin{equation*}
\bar{w}=\frac{1}{L^{2}}\int_{0}^{L/2}\frac{w_{0}}{1+\gamma r^{2}}2\pi rdr=%
\frac{\pi w_{0}}{L^{2}\gamma }\ln \left[ 1+\gamma \left( \frac{L}{2}\right)
^{2}\right] .
\end{equation*}%
From the comparison of the coefficients in \ front of similar terms in Eq. (%
\ref{substitution}), we obtain
\begin{equation}
\gamma =\frac{\exp \left( L^{2}/\pi \right) -1}{(L/2)^{2}},\;\bar{w}=\frac{%
\pi W_{0}\ln \left[ 1+\gamma (L/2)^{2}\right] }{\gamma L^{2}},\;k=(1/8)(\bar{%
w}-1)-\frac{Q}{4},  \label{coeff}
\end{equation}%
and amplitude $w_{0}$ of ansatz (\ref{Lorentz}) is determined by the
normalization condition,
\begin{equation*}
\int \int w^{2}\left( x,y\right) dxdy\simeq \int_{0}^{L/2}\frac{w_{0}^{2}}{%
\left( 1+\gamma r^{2}\right) ^{2}}2\pi rdr=\frac{\pi \left( L/2\right) ^{2}}{%
1+\gamma (L/2)^{2}}w_{0}^{2}=\frac{P}{4},
\end{equation*}%
which yields%
\begin{equation}
w_{0}=\frac{1}{L}\sqrt{\frac{P}{\pi }\left[ 1+\gamma (L/2)^{2}\right] }.
\label{W0}
\end{equation}

According to Eq. (\ref{coeff}), width $1/\sqrt{\gamma }$ of Lorentzian (\ref%
{Lorentz}) strongly (exponentially) depends on the system's size $L$, due to
the strong nonlocality of Eq. (\ref{linear-w}) in the limit of $\alpha
\rightarrow 0$. For $L\rightarrow \infty $, the power density corresponding
to the Lorentzian may be roughly approximated by the 2D delta-function:%
\begin{equation*}
4w^{2}(x,y)\simeq P\delta (x)\delta (y).
\end{equation*}

\begin{figure}[h]
\begin{center}
\includegraphics[height=4.cm]{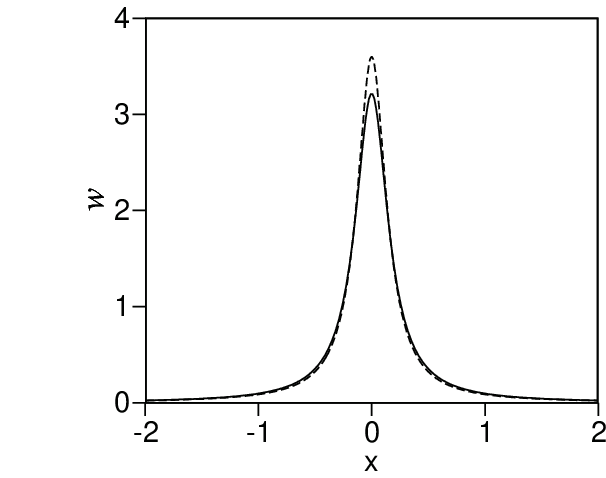}
\end{center}
\caption{The profile of the GS single-color state $w$ produced by the linear
Schr\"{o}dinger equation Eq.~(\protect\ref{linear-w}) at $\protect\alpha %
\rightarrow 0$ and $P=4$, in the domain of size $L\times L=4\times 4$. The
solid line is the numerical result produced by the IT method, and the dashed
line is the Lorentzian ansatz (\protect\ref{Lorentz}) with $\protect\gamma %
=40.47$ and $w_{0}=3.60$, as given by Eqs. (\protect\ref{coeff}) and (%
\protect\ref{W0}).}
\label{fig8a}
\end{figure}

Figure \ref{fig8a} shows the cross sections, drawn through $y=0$, of the
numerically found solution $w(x,y)$ (the solid line) for $P=4$, and of the
corresponding approximate solution (\ref{Lorentz}), with $\gamma =40.47$ and
$W_{0}=3.60$, as given by Eqs. (\ref{coeff}) and (\ref{W0}). It is seen that
the Lorentzian ansatz yields an accurate approximation in this case.

\paragraph{The single-color states with finite LI, $0<\protect\alpha <2$}

A nontrivial problem is the stability of the single-color solutions of the
linear equation (\ref{linear-w}) against parametric excitations, i.e., small
perturbations of the FF field $U$ driven by term $WU^{\ast }$ in Eq. (\ref%
{potU}). In the case of the non-fractional diffraction ($\alpha =2$), the
parametric instability of such states, with $m_{2}=0$, $1$, and $2$, was
investigated in Ref. \cite{Hidetsugu}. It was found that the single-color
states with each value of $m_{2}$ are stable for values of the power below a
certain threshold,
\begin{equation}
P<\left( P_{\mathrm{thr}}\right) _{m}^{(\mathrm{single-color})},
\label{Psingle}
\end{equation}%
which depends on the mismatch. At the threshold, the single-color states
with even $m_{2}$ (actually, $m_{2}=0$ and $2$ were considered) underwent a
pitchfork bifurcation \cite{bif}, which gave rise to stable two-color states
with FF vorticity $m_{2}/2$ at values of the power above the threshold. The
single-color mode with $m_{2}=1$, which cannot bifurcate into a two-color
vortex, was spontaneously transformed, above the threshold, into an
apparently unstable state with chaotic oscillations.

For a crude estimate of the onset of the parametric instability of the
single-color states, with $U=0$, we first analyze this issue in the case of
the uniform state (in the absence of the trapping potential), which is
represented by an evident solution of Eq. (\ref{W}),
\begin{equation}
W(z)=W_{0}\exp \left( -iQz/2\right) ,  \label{WW}
\end{equation}%
with a constant real amplitude $W_{0}$. Then, a small FF perturbation in
substituted in Eq. (\ref{U}) as $U\left( x,y,z\right) =U_{0}(z)\exp
(ik_{x}x+ik_{y}y)$, with amplitude $U_{0}(z)$ satisfying the equation%
\begin{equation}
i\frac{dU_{0}}{dz}=\frac{1}{2}k^{\alpha }U_{0}+W(z)U_{0},  \label{d/dz}
\end{equation}%
where $k=\sqrt{k_{x}^{2}+k_{y}^{2}}$. Inserting $W(z)$ from Eq. (\ref{WW})
in Eq. (\ref{d/dz}), the ensuing solution for the FF perturbation amplitude
is%
\begin{equation}
U_{0}=U_{0}^{(0)}\exp \left[ i\left( Q/4+i\lambda (k)\right) z\right] ,
\label{U00}
\end{equation}%
with eigenvalue
\begin{equation}
\lambda (k)=\sqrt{\left( Q/4-k^{\alpha }\right) ^{2}-W_{0}^{2}}.
\label{lambda}
\end{equation}

For $Q=-1$, Eq. (\ref{U00}) yields real $\lambda $, i.e., stability, for all
values of the perturbation wavenumber $k$, provided that the amplitude of
the uniform SH state is small enough, $W_{0}<1/4$, irrespective of the LI
value $\alpha $. Otherwise, as well as for all values of $W_{0}$ in the case
of $Q=0$ and $Q=+1$, Eq. (\ref{lambda}) yields imaginary $\lambda $, i.e.,
the parametric instability, at some values of $k$.

\begin{figure}[h]
\begin{center}
\includegraphics[height=4.cm]{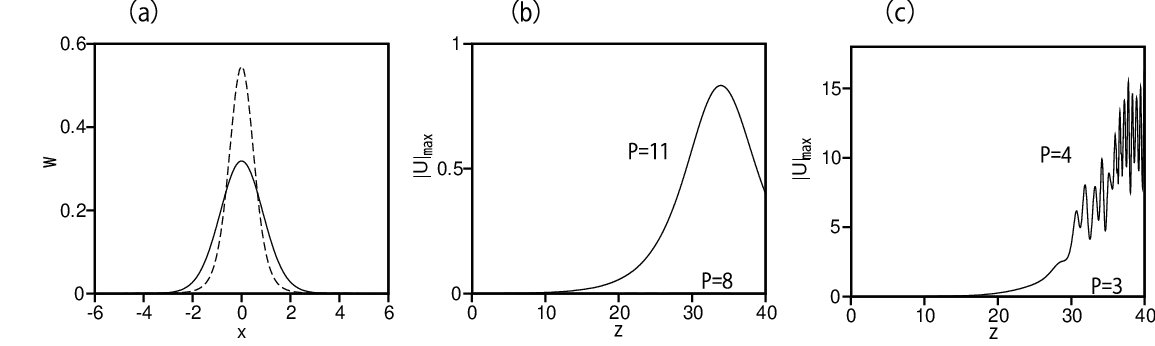}
\end{center}
\caption{(a) Profiles of the trapped single-color GS, $w(x)$ (with $m_{2}=0$
and $P=1$), are shown in cross section $y=0$, as produced by the numerical
IT solution of Eqs. (\protect\ref{potU}) and (\protect\ref{potW}) with $%
\protect\alpha =1.5$\ and $0.5$ (the solid and dashed lines, respectively).
(b) The evolution of the amplitude, $|U(x=0)|$, for the same GS solutions
with powers $P=8$ (stable) and $11$ (unstable), at $\protect\alpha =1.5$.
(c) The same as in (b), but for $P=3$ (stable) and $4$ (unstable) at $%
\protect\alpha =0.5$. In all cases, $Q=-1$.}
\label{fig8}
\end{figure}

\begin{figure}[h]
\begin{center}
\includegraphics[height=4.cm]{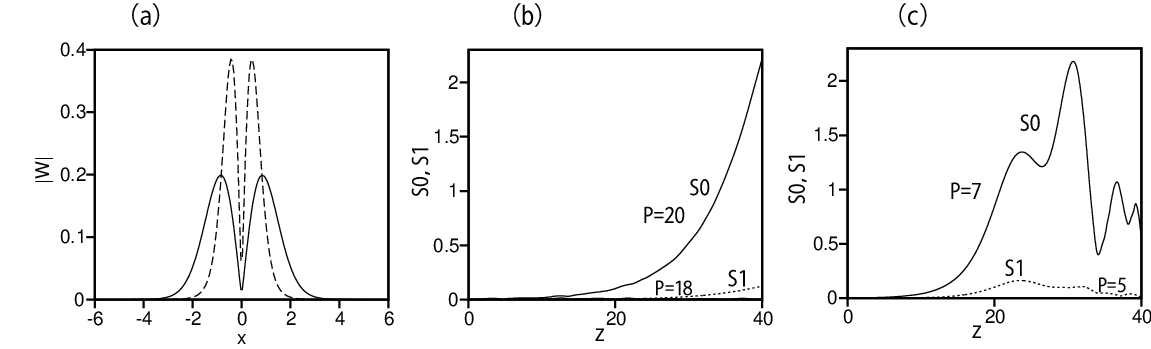}
\end{center}
\caption{(a) Profiles of the trapped single-color state, $|W(x)|$. with
vorticity $m_{2}=1$ and $P=1$ are shown in cross section $y=0$, as produced
by the numerical solution of Eqs. (\protect\ref{potU}) and (\protect\ref%
{potW}) with $\protect\alpha =1.5$\ and $0.5$ (the solid and dashed lines,
respectively). (b) The evolution of form-factors (\protect\ref{S0S1}) for
the vortex states with $m_{2}=1$ and powers $P=18$ (stable) and $20$
(unstable), the LI being $\protect\alpha =1.5$. (c) The same as in (b), but
for powers $P=5$ (stable) and $7$ (unstable), and LI $\protect\alpha =0.5$.}
\label{fig9}
\end{figure}

Proceeding to the single-color states trapped in the HO potential, which are
produced as stationary solutions of the linear equation (\ref{linear}),
examples of the GS with zero vorticity and excited state with vorticity $%
m_{2}=1$ are plotted, for LI values $\alpha =1.5$ and $0.5$, in Figs. \ref%
{fig8}(a) and \ref{fig9}(a), respectively. Here, $Q=-1$ is taken, as it
facilitates the stability of the single-color states, pursuant to Eq. (\ref%
{lambda}). The fractional diffraction is weaker for smaller LI $\alpha $,
hence the respective profiles are sharper in Figs. \ref{fig8}(a) and \ref%
{fig9}(a).

The parametric (in)stability of the single-color ($W$-only) states was
tested by seeding a perturbation in the $U$ component, with small amplitude $%
\varepsilon $ and the shape naturally chosen, at $z=0$, as
\begin{equation}
U_{m_{2}=0}^{(\mathrm{pert})}\left( x,y\right) =\varepsilon \exp \left(
-r^{2}\right) ,U_{m_{2}=1}^{(\mathrm{pert})}\left( x,y\right) =\varepsilon
\left( 1+re^{i\theta }\right) \exp \left( -r^{2}\right) ,U_{m_{2}=2}^{(%
\mathrm{pert})}\left( x,y\right) =\varepsilon re^{i\theta }\exp \left(
-r^{2}\right) ,  \label{pert}
\end{equation}%
and running direct simulations of the perturbed evolution of the
single-color states with the indicated values of the SH vorticity $m_{2}$.
The results, which are displayed in Figs. \ref{fig8}(b,c) and \ref{fig9}%
(b,c) for $m_{2}=0$ (the GS) and $m_{2}=1$, respectively, demonstrate the
existence of a critical (threshold) power of the single-color states, $%
\left( P_{\mathrm{thr}}\right) _{m_{2}}^{(\mathrm{single-color})}(\alpha )$
(cf. Eq. (\ref{Psingle})), such that the single-color states are stable with
the power below the critical value, and unstable above it. In particular,
for the vortex states with $m_{2}=1$ the development of the instability is
illustrated by the evolution of form-factors%
\begin{equation}
S_{0}=\int \int \left\vert U\left( x,y\right) \right\vert dxdy,~S_{1}=\int
\int \left\vert U\left( x,y.\right) e^{-i\theta }\right\vert dxdy.
\label{S0S1}
\end{equation}

The numerical findings yield the following values of the critical
(threshold) power below which the single-color states keep their stability:%
\begin{gather}
\left( P_{\mathrm{thr}}\right) _{m_{2}=0}^{(\mathrm{single-color})}\left(
\alpha =0.1\right) \approx 0.35,~\left( P_{\mathrm{thr}}\right) _{m_{2}=0}^{(%
\mathrm{single-color})}\left( \alpha =0.5\right) \approx 3.5,~\left( P_{%
\mathrm{thr}}\right) _{m_{2}=0}^{(\mathrm{single-color})}\left( \alpha
=1.5\right) \approx 9.3;  \notag \\
\left( P_{\mathrm{thr}}\right) _{m_{2}=1}^{(\mathrm{single-color})}\left(
\alpha =0.5\right) \approx 5.9,~\left( P_{\mathrm{thr}}\right) _{m_{2}=1}^{(%
\mathrm{single-color})}\left( \alpha =1.5\right) \approx 18.6;  \notag \\
\left( P_{\mathrm{thr}}\right) _{m_{2}=2}^{(\mathrm{single-color})}\left(
\alpha =0.5\right) \approx 7.55,~\left( P_{\mathrm{thr}}\right) _{m_{2}=2}^{(%
\mathrm{single-color})}\left( \alpha =1.5\right) \approx 23.9.  \label{thr}
\end{gather}%
Note that the trapping potential makes it possible to create the stable
single-color modes even at $\alpha \leq 1$, cf. the free-space stability
interval (\ref{no-coll}). The threshold value increases with the increase of
$\alpha $ (stronger diffraction, corresponding to larger $\alpha $, makes it
easier to suppress the growth of small perturbations), and is larger for
larger vorticity $m_{2}$.

\subsubsection{Two-color states}

\begin{figure}[h]
\begin{center}
\includegraphics[height=4.cm]{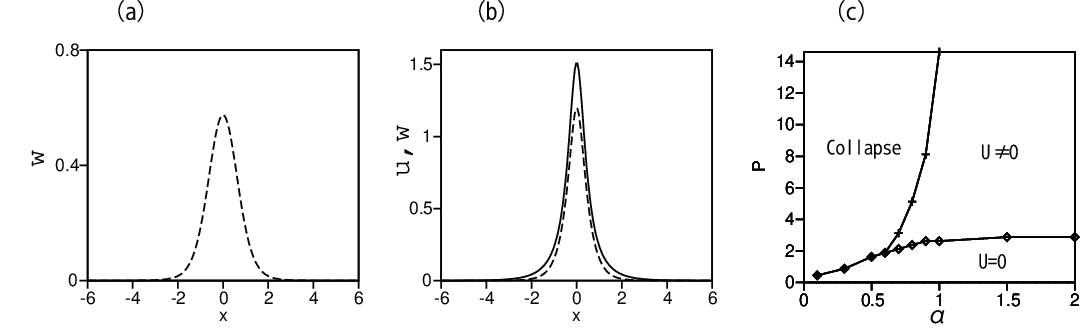}
\end{center}
\caption{(a) The cross section of the stable single-color GS, represented
solely by the SH component, $w$ (with $m_{2}=0$, and $U=0$), as produced by
the IT method applied to Eqs. (\protect\ref{potU}) and (\protect\ref{potW})
for $\protect\alpha =0.8$, $Q=0$, and $P=1.8$. (b) The stable two-color GS,
which includes both the FF and SH compomnents ($u$ and $w$, respectively),
obtained for $\protect\alpha =0.8$, $Q=0$, and $P=4.5$. (c) The GS phase
diagram in the parameter plane $\left( \protect\alpha ,P\right) $. Symbols $%
\mathrm{U=0}$ and $\mathrm{U}\mathrm{\neq 0}$ designate areas which are
populated, respectively, by the stable single- and two-color GSs. In the
limit of $\protect\alpha =2$, the stability diagram carries over into the
one known in the system with the normal (non-fractional) diffraction
\protect\cite{Hidetsugu}.}
\label{fig10}
\end{figure}

The destabilization of the single-color modes with $m_{2}=0$ and $2$ at the
threshold points identified as per Eq. (\ref{thr}) is explained by a
bifurcation which gives rise to two-color states, with $U\neq 0$. For the
GS, with $m_{2}=0$, the transition from the single-color state to the
two-color one, in the system with $Q=0$, is illustrated by panels (a) and
(b) in Fig. \ref{fig10}. Systematically collected numerical results are
presented in Fig. \ref{fig10}(c) as the chart in the (LI, power) plane,
which includes areas populated by stable single- and two-color GSs, as well
as the area (at larger $P$ and smaller $\alpha $) where the system blows up
(the collapse takes place). We stress that the trapped GSs of both the
single- and two-color types may be stable at $\alpha \leq 1$, while, as said
above, all the stationary free-space states are unstable $\alpha \leq 1$.
Note that the threshold value of the power at the single-color GS boundary
is going to vanish, $\left( P_{\mathrm{thr}}\right) _{m_{2}=0}\rightarrow 0$
at $\alpha \rightarrow 0$. This feature complies with a very small value, $%
\left( P_{\mathrm{thr}}\right) _{m_{2}=0}^{(\mathrm{single-color})}\left(
\alpha =0.1\right) \approx 0.35$ at small $\alpha =0.1$ and $Q=-1$ in Eq. (%
\ref{thr}) (which is different from $Q=0$ corresponding to Fig. \ref{fig10}%
(c)).
\begin{figure}[h]
\begin{center}
\includegraphics[height=4.cm]{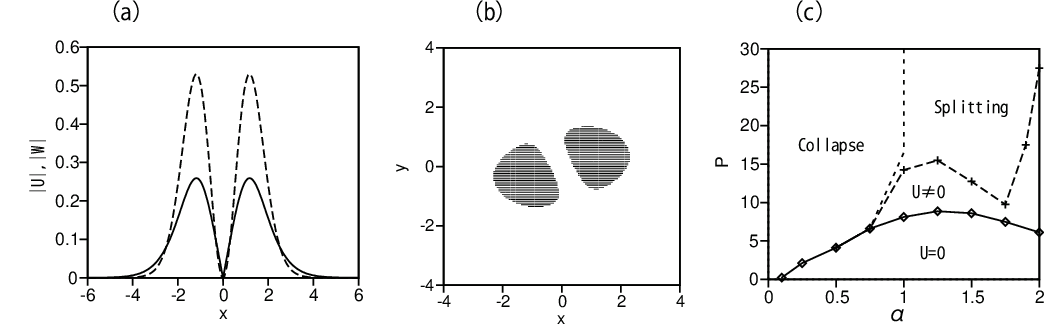}
\end{center}
\caption{(a) A stable stationary two-color vortex state with $m_{2}=2$, as
produced by the IT simulation method applied to Eqs. (\protect\ref{potU})
and (\protect\ref{potW}) with $\protect\alpha =1.5$, $Q=0$, and power $P=10$%
. $|U|$ and $|W|$ are plotted by the solid and dashed lines, respectively.
(b) A snapshot of the rotating pattern of $|U\left( x,y\right) |>0.5$ at $%
\protect\alpha =1.5$, $Q=0$, and $P=24$, as produced by the spontaneous
splitting (fission) of the unstable vortex state. (c) The stability diagram
in the parameter plane of $\left( \protect\alpha ,P\right) $. Symbols $%
\mathrm{U=0}$ and $\mathrm{U}\mathrm{\neq 0}$ designate areas which are
populated, respectively, by the stable single- and two-color vortex states.
The \textrm{splitting} domain is populated by rotating dipoles, such as the
one shown in panel (b). Similar to the situation plotted in Fig. \protect\ref%
{fig10}, in the limit of $\protect\alpha =2$, the stability diagram carries
over into the one known in the system with the normal (non-fractional)
diffraction \protect\cite{Hidetsugu}.}
\label{fig11}
\end{figure}

A similar picture for the vortex states, with $m_{2}=2$, is presented in
Fig. \ref{fig11}. Cross sections of the FF and SH components of a stable
two-color state with $m_{1}=1$ and $m_{2}=2$ are plotted in Fig. \ref{fig11}%
(a) for $\alpha =1.5$, $Q=0$, and $P=10$ (examples of stable single-color
vortex states with $m_{2}=2$ are not shown here). For $\alpha =1.5$ and $Q=0$%
, the vortex state with $m_{2}=2$ is unstable against spontaneous splitting
in two fragments at $P>12.75$. Obeying the conservation of the angular
momentum (see Eq. (\ref{M})), the fragments build a rotating dipole state,
as shown in Fig. \ref{fig11}(b) for $P=24$. The angular velocity of the
rotation of the emerging state is $\approx 0.34$.

Finally, the phase diagram for the vortex states is plotted in Fig. \ref%
{fig11}(c) in the plane of LI and power. Similar to the stability chart for
the GS, displayed in Fig. \ref{fig10}(c), labels $\mathrm{U=0}$ and $\mathrm{%
U}\mathrm{\neq 0}$ designate parametric areas populated by the single- and
two-color states, respectively. Unlike the GS chart, where the entire
instability area is located, in Fig. \ref{fig10}(c), at $\alpha <1$ and,
accordingly, represents the collapse, the instability area in Fig. \ref%
{fig11}(c) is separated by the vertical dashed line into the collapse domain
at $\alpha <1$, and in the splitting one at $\alpha >1$.

\section{Conclusion}

\label{Sec V}

We have introduced the model for the spatial-domain copropagation of the FF
(fundamental-frequency) and SH (second-harmonic) waves in the bulk waveguide
with the $\chi ^{(2)}$ nonlinearity and effective fractional transverse
diffraction acting onto both components. The main objective of the analysis
is to construct families of 2D fundamental (zero-vorticity) and vortical
solitons and explore their stability. In the free-space system, we have
obtained a family of stable fundamental solitons, in agreement with the VK
(Vakhitov-Kolokolov) criterion, while all vortex solitons are unstable
against spontaneous fission. Because the fractional diffraction breaks the
system's Galilean invariance, we have also produced solutions for moving
stable fundamental solitons, and demonstrated that collisions between them
are destructive.

Different results are produced for the FF-SH fractional $\chi ^{(2)}$ system
including the trapping HO (harmonic-oscillator) potential. The trap admits
the existence of both single-color (SH-only) and two-color (FF-SH) solitons,
including the fundamental ones and solitons carrying the embedded vorticity.
The increase of the soliton's power drives the bifurcation which transforms
the single-color solitons into their two-color counterparts. The stability
areas have been identified for the solitons of both types, with zero and
nonzero vorticities alike.

As extension of the present analysis, it may be interesting to introduce the
2D three-wave system which, in terms of optics, corresponds to the
copropagation of two different FF polarization components and the single SH
wave, coupled by the $\chi ^{(2)}$ interaction (the three-wave system is
also known as the one realizing the Type-II $\chi ^{(2)}$ interaction \cite%
{Buryak}). In that case, an additional control parameter is the
birefringence of the FF components. The structure and stability of the
solitons in the three-wave system may be essentially different.

\section*{Acknowledgments}

The work of B.A.M. was supported, in part, by the Israel Science Foundation
through Grant No. 1695/22.

\end{document}